\documentclass[aps,twocolumn,showpacs,groupedaddress]{revtex4-1}
\usepackage{amssymb}
\usepackage{mathrsfs}
\usepackage{pstricks}
\usepackage{graphicx}
\usepackage{hyphenat}
\usepackage{hyperref}

\begin{document}


\title{\bf Renormalization group approach for Lorentz\hyp{}violating scalar field theory at all loop orders}


\author{William C. Vieira}
\email{william.vieira19@gmail.com}
\affiliation{\it Departamento de F\'\i sica, Universidade Federal do Piau\'\i, 64049-550, Teresina, PI, Brazil}
\author{Paulo R. S. Carvalho}
\email{prscarvalho@ufpi.edu.br}
\affiliation{\it Departamento de F\'\i sica, Universidade Federal do Piau\'\i, 64049-550, Teresina, PI, Brazil}




\begin{abstract}
We compute, both explicitly up to next\hyp{}to\hyp{}leading order and in a proof by induction for all loop levels, the critical exponents for thermal Lorentz\hyp{}violating O($N$) self\hyp{}interacting scalar field theory. They are evaluated in a massless theory renormalized at arbitrary external momenta, where a reduced number of Feynman diagrams is needed. The results are presented and shown to be identical to that found previously in distinct theories renormalized at different renormalization schemes. At the end we give both mathematical explanation and physical interpretation for them based on coordinates redefinition techniques and symmetry ideas, respectively.
\end{abstract}


\maketitle


\section{Introduction}\label{Introduction} 

\par The renormalization group technique was designed for solving some non\hyp{}trivial problems: that involving many length or equivalently energy scales \cite{PhysRevB.4.3174,PhysRevB.4.3184}. These problems can be understood only if we consider a theory that takes into account many coupled degrees of freedom. We can cite the high energy physics effects of deep inelastic scattering and electron\hyp{}positron annihilation as examples of such problems \cite{DJGross}. On the other hand, we may also mention the task of computation of the critical exponents for a thermal continuous transition in the theory of phase transitions and critical phenomena in the low energy physics realm \cite{PhysRevLett.28.240,PhysRevLett.28.548,Wilson197475}. In this work, we consider some symmetry aspects of the latter problem.  

\par The precise determination of the critical exponents for systems undergoing a thermal continuous phase transition attained its state of the art when perturbative higher\hyp{}loop field\hyp{}theoretic renormalization group and $\epsilon$\hyp{}expansion tools were applied . The loop quantum corrections to classical Landau critical exponents were obtained up to five\hyp{}loop order \cite{Kleinert1993545,Kleinert}. These quantum corrections have their physical origin in the fluctuations of a quantum field. This field acts as an order parameter, which indicates a transition when occurs its symmetry breaking, whose mean value represents for example the magnetization in magnetic systems. As one has to treat a fluctuating physical quantity i.e. the fluctuating field, statistic mechanical ideas in the canonical ensemble formalism at a finite temperature $T$ were employed, where the exponentiation of a $d$\hyp{}dimensional Euclidean Lagrangian density played the role of a Boltzmann factor. This is why such an approach of statistical mechanics is named thermal statistical field theory \cite{LeBellac}. The Lagrangian density, known as Landau\hyp{}Ginzburg Lagrangian density, contains the complete information needed for an accurate description of the required critical properties. This Lagrangian density is a function of the $N$\hyp{}component field and its derivatives and it is constructed based on certain constrains. Among these constrains are the maintenance of the symmetry of the order parameter and the inclusion of operators of the field belonging to a class of operators called relevant operators. For the systems studied here the Lagrangian density is composed of both Lorentz\hyp{}invariant (LI) and O($N$) (the $N$\hyp{}dimensional orthogonal group) relevant operators. A relevant operator is such that its canonical dimension, inferred by power counting analysis, is less than or equal to four \cite{Amit}. We could consider operators with canonical dimensions greater than four, the high composite operators, but they would provide negligible corrections to the results. These corrections are called corrections to irrelevant operators \cite{BrezinZinnJustin9483,BrezinZinnJustin10849,PhysRevD.7.2927}. Once we have defined the Lagrangian density, the critical exponents can be extracted from the non\hyp{}trivial scaling properties of the renormalized primitively one\hyp{}particle irreducible ($1$PI) vertex parts with amputated external legs, namely the $\Gamma_{R}^{(2)}$ and $\Gamma_{R}^{(4)}$ functions \cite{Amit}. These functions are finite as opposed to their corresponding infinite or bare counterparts $\Gamma_{B}^{(2)}$ and $\Gamma_{B}^{(4)}$, respectively. Their infinities are absorbed multiplicatively by renormalization constants. The critical exponents are then directly associated to the anomalous dimensions of the $1$PI vertex parts near the critical point, i.e. the anomalous dimensions of the field and composite field. They are associated to the exponents $\eta$ and $\nu$, respectively. The four remaining critical exponents $\beta$, $\gamma$, $\alpha$ and $\delta$ needed for a complete characterization of the critical behavior of the system can be obtained by using four independent scaling relations among them. Massles and massive theories correspond to critical (at the critical temperature) and non\hyp{}critical (not at but near to the critical temperature) theories, respectively. The universality of the critical exponents means that they are identical when evaluated in massless or massive theories, although the theories be renormalized at distinct renormalization schemes, i.e. they present different renormalization constants, $\beta$\hyp{}function, anomalous dimensions, fixed points etc in intermediate steps. Following the program above, we obtain in this work the critical exponents for massless thermal Lorentz\hyp{}violating (LV) O($N$) self\hyp{}interacting $\lambda\phi^{4}$ scalar field theory in the minimal subtraction scheme. This scheme is characterized by its simplicity because a reduced number of diagrams is needed, elegance and generality since all the Feynman diagrams are evaluated at arbitrary external momenta, as opposed to the previously applied normalization condition renormalization scheme with the respective diagrams computed at fixed nonvanishing external momenta. The latter and the BPHZ (were a larger number of diagrams and counterterms has to be computed) renormalization schemes were applied to renormalize massless and massive theories \cite{EurophysLett10821001}, whose results are to be compared with the ones found independently here. To the usual $\lambda\phi^{4}$ scalar field theory we add the only four dimensional LV O($N$) relevant operator $K_{\mu\nu}\partial^{\mu}\phi\partial^{\nu}\phi$ with dimensionless constant coefficients $K_{\mu\nu}$  \cite{PhysRevD.84.065030}. It plays a role of a constant background field. It is symmetric ($K_{\mu\nu} = K_{\nu\mu}$) and equal for all $N$ components of the field which preserves the O($N$) symmetry of the $N$\hyp{}component field. We have a slightly violation of the Lorentz symmetry when the coefficients do not transform as a second order tensor under Lorentz transformations, i.e. $K_{\mu\nu}^{\prime} \neq \Lambda_{\mu}^{\rho}\Lambda_{\nu}^{\sigma} K_{\rho\sigma}$ and $|K_{\mu\nu}|\ll 1$. We could have introduced higher derivative operators but they would contribute with negligible corrections as irrelevant operators. There were some works analyzing particular combinations of these coefficients, specifically just their traceless part \cite{PhysRevD.7.2911}. So the investigation in this paper is the most general in this respect and furthermore makes a clear identification of the effect of a LV symmetry breaking relevant operator on the outcome for the critical exponents. As the critical exponents are universal, they are identical for realistic quite different systems as a fluid and a ferromagnet. They depend only on the dimension $d$ of the system, $N$ and symmetry of some $N$\hyp{}component order parameter(a mean value of the field, so this symmetry is directly related to the symmetry of the field), and if the interactions present are of short\hyp{} or long\hyp{}range type. When two or more systems have their critical properties described by an equal set of critical exponents, these systems belong to the same universality class. A general universality class, the O($N$) universality class treated here, encompasses the particular models: Ising ($N=1$), XY ($N=2$), Heisenberg ($N=3$), self\hyp{}avoiding random walk ($N=0$) and spherical ($N \rightarrow \infty$) for short\hyp{}range interactions \cite{Pelissetto2002549}. Recently there were many studies on the dependence of the critical exponents on the more obvious parameters as $d$ \cite{PhysRevB.86.155112,PhysRevE.71.046112} and $N$ \cite{PhysRevLett.110.141601,Butti2005527,PhysRevB.54.7177} as well as on the less intuitive symmetry of the order parameter \cite{PhysRevE.78.061124,Trugenberger2005509}. Studying the effect of a symmetry breaking mechanism on the values for the critical exponents is the aim of this work.            

\par This paper is organized as follows. In Sect. \ref{Minimal subtraction scheme for the critical theory} we obtain, explicitly up to next\hyp{}to\hyp{}leading order, the critical exponents for the critical LV O($N$) self\hyp{}interacting $\lambda\phi^{4}$ scalar field theory in the minimal subtraction scheme, where the massless theory is renormalized at arbitrary nonvanishing external momenta by computing just a few Feynman diagrams. In Sect. \ref{All-loop critical exponents in the minimal subtraction scheme} we generalize the outcomes for the critical exponents for an arbitrary loop order. In Sect. \ref{Conclusions} we present our conclusions.

\section{Minimal subtraction scheme for the critical theory}\label{Minimal subtraction scheme for the critical theory}

\par The bare $d$\hyp{}dimensional Euclidean Lagrangian density for the critical theory is given by 
\begin{eqnarray}\label{huytrji}
&&\mathcal{L}_{0} = \frac{1}{2}\partial^{\mu}\phi_{0}\partial_{\mu}\phi_{0} + \frac{1}{2}K_{\mu\nu}\partial^{\mu}\phi_{0}\partial^{\nu}\phi_{0} + \frac{\lambda_{0}}{4!}\phi_{0}^{4} \nonumber \\ && + \frac{1}{2}t_{0}\phi_{0}^{2}
\end{eqnarray} 
The second term above beaks the Lorentz symmetry and $\phi_{0}$ and $\lambda_{0}$ are the bare field and coupling constant, respectively. As in a critical theory we have to renormalize the bare primitively divergent vertex functions, $\Gamma_{B}^{(2)}$, $\Gamma_{B}^{(4)}$ and $\Gamma_{B}^{(2,1)}$ (for composite fields), the last term in Eq. \ref{huytrji} generates the composite field $1$PI vertex parts. We follow the steps and conventions of the Ref. \cite{Amit} inspired on the original work \cite{BrezinLeGuillouZinnJustin,ZinnJustin}. These functions are renormalized multiplicatively 
\begin{eqnarray}\label{uhygtfrd}
\Gamma_{R}^{(n, l)}(P_{i}, Q_{j}, g, \kappa) = Z_{\phi}^{n/2}Z_{\phi^{2}}^{l}\Gamma_{B}^{(n, l)}(P_{i}, Q_{j}, \lambda_{0}, \kappa).
\end{eqnarray}
where the divergences, in the form of poles, are extracted minimally in the minimal subtraction scheme and absorbed in the renormalization constants for the field $Z_{\phi}$ and composite field $Z_{\phi^{2}}$, respectively. The external momenta $P_{i}$ ($i = 1, \cdots, n$) and $Q_{j}$ ($j = 1, \cdots, l$) are associated to the field and composite field external legs insertions, respectively. The $g$ parameter is the renormalized coupling constant. It can be written in terms of a dimensionless one by $g = u\kappa^{\epsilon}$ (for the unrenormalized coupling constant, $\lambda_{0} = u_{0}\kappa^{\epsilon}$), where $\kappa$ is an arbitrary momentum scale and $\epsilon = 4 - d$ is a small parameter for avoiding the divergent point $d = 4$ in the $1$PI vertex parts. The dimensional $\kappa$ parameter will be used for redefining all the diagrams momenta as dimensionless quantities through $P^{\mu} = \kappa \widehat{P}^{\mu}$, where $\widehat{P}^{\mu}$ is dimensionless. Thus making another redefinition, i.e. $\widehat{P}^{\mu} \rightarrow P^{\mu}$ means that $P^{\mu}$ is dimensionless and that the dependence on $\kappa$ of diagrams will be absorbed on the coupling constant. We can then expand the bare dimensionless coupling and renormalization constants as
\begin{eqnarray}\label{suhufjifjvf}
u_{0} = u\left( 1 + \sum_{i=1}^{\infty} a_{i}(\epsilon)u^{i}\right),
\end{eqnarray}
\begin{eqnarray}
Z_{\phi} = 1 + \sum_{i=1}^{\infty} b_{i}(\epsilon)u^{i},
\end{eqnarray}
\begin{eqnarray}\label{iaifkdvkvkck}
\overline{Z}_{\phi^{2}} = 1 + \sum_{i=1}^{\infty} c_{i}(\epsilon)u^{i},
\end{eqnarray}
where it will be more convenient to use the renormalization constant $\overline{Z}_{\phi^{2}} \equiv Z_{\phi}Z_{\phi^{2}}$ instead of $Z_{\phi^{2}}$, for renormalization of $\Gamma^{(2,1)}$ purposes. For a single-component field the bare $1$PI vertex parts to be renormalized, up to next\hyp{}to\hyp{}leading order, can be written as
\begin{eqnarray}\label{gtfrdrdes}
\Gamma^{(2)}_{B}(P, u_{0}, \kappa) = P^{2}( 1 - B_{2}u_{0}^{2} + B_{3}u_{0}^{3}),
\end{eqnarray}
\begin{eqnarray}
\Gamma^{(4)}_{B}(P_{i}, u_{0}, \kappa) = \kappa^{\epsilon}u_{0}[ 1 - A_{1}u_{0} + (A_{2}^{(1)} + A_{2}^{(2)})u_{0}^{2}],
\end{eqnarray}
\begin{eqnarray}\label{gtfrdesuuji}
\Gamma^{(2,1)}_{B}(P_{1}, P_{2}, Q_{3}, u_{0}, \kappa) = 1 - C_{1}u_{0} + (C_{2}^{(1)} + C_{2}^{(2)})u_{0}^{2}. \nonumber \\
\end{eqnarray}
In the last Eq. we have $Q_{3} = -(P_{1} + P_{2})$. The coefficients have the form
\begin{eqnarray}
A_{1} = \frac{(N + 8)}{18}\left[ \parbox{10mm}{\includegraphics[scale=1.0]{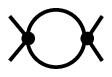}} + 2 \hspace{2mm} \text{perm.} \right]
\end{eqnarray}
\begin{eqnarray}
A_{2}^{(1)} = \frac{(N^{2} + 6N + 20)}{108}\left[ \parbox{10mm}{\includegraphics[scale=1.0]{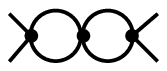}}\quad\quad  + 2 \hspace{2mm} \text{perm.} \right],
\end{eqnarray}
\begin{eqnarray}
A_{2}^{(2)} = \frac{(5N + 22)}{54}\left[ \parbox{10mm}{\includegraphics[scale=1.0]{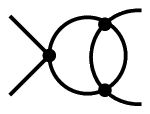}}\quad+ 5 \hspace{2mm} \text{perm.} \right],
\end{eqnarray}
\begin{eqnarray}
B_{2} = \frac{(N + 2)}{18}\hspace{1mm}\parbox{12mm}{\includegraphics[scale=1.0]{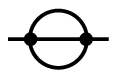}},
\end{eqnarray}
\begin{eqnarray}
B_{3} = \frac{(N + 2)(N + 8)}{108}\hspace{1mm}\parbox{12mm}{\includegraphics[scale=1.0]{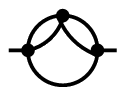}},
\end{eqnarray}
\begin{eqnarray}
C_{1} = \frac{(N + 2)}{6}\hspace{1mm}\parbox{10mm}{\includegraphics[scale=1.0]{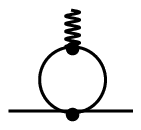}}\hspace{4mm}, 
\end{eqnarray}
\begin{eqnarray}
C_{2}^{(1)} = \frac{(N + 2)^{2}}{36}\hspace{1mm}\parbox{10mm}{\includegraphics[scale=1.0]{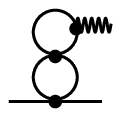}}\hspace{2mm},
\end{eqnarray}
\begin{eqnarray}
C_{2}^{(2)} = \frac{(N + 2)}{6}\hspace{1mm}\parbox{10mm}{\includegraphics[scale=1.0]{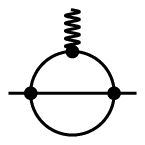}}\hspace{5mm}.
\end{eqnarray}

\par The renormalized theory is attained if we demand that the dimensional poles be minimally subtracted and absorbed in the renormalization constants. The finite vertex parts then satisfy the renormalization group equation 
\begin{eqnarray}\left( \kappa\frac{\partial}{\partial\kappa} + \beta\frac{\partial}{\partial u} - \frac{1}{2}n\gamma_{\phi} + l\gamma_{\phi^{2}} \right)\Gamma_{R}^{(n, l)} = 0
\end{eqnarray}
where 
\begin{eqnarray}\label{kjjffxdzs}
\beta(u) = \kappa\frac{\partial u}{\partial \kappa} = -\epsilon\left(\frac{\partial\ln u_{0}}{\partial u}\right)^{-1},
\end{eqnarray}
\begin{eqnarray}\label{koiuhygtf}
\gamma_{\phi}(u) = \beta(u)\frac{\partial\ln Z_{\phi}}{\partial u},
\end{eqnarray}
\begin{eqnarray}
\gamma_{\phi^{2}}(u) = -\beta(u)\frac{\partial\ln Z_{\phi^{2}}}{\partial u}.
\end{eqnarray}
We will be concerned with the function
\begin{eqnarray}\label{udgygeykoiuhygtf}
\overline{\gamma}_{\phi^{2}}(u) = -\beta(u)\frac{\partial\ln \overline{Z}_{\phi^{2}}}{\partial u} \equiv \gamma_{\phi^{2}}(u) - \gamma_{\phi}(u)
\end{eqnarray}
instead of $\gamma_{\phi^{2}}(u)$, for convenience. The $\beta$\hyp{}function and the anomalous dimensions in the Eqs. \ref{kjjffxdzs}, \ref{koiuhygtf} and \ref{udgygeykoiuhygtf} can be expressed as functions of the coefficients $a_{1}$, $\cdots$, $c_{2}$ through the Eqs. \ref{suhufjifjvf}\hyp{}\ref{iaifkdvkvkck}, up to next\hyp{}to\hyp{}leading level, as 
\begin{eqnarray}
\beta(u) = -\epsilon u[ 1 - a_{1}u + 2(a_{1}^{2} - a_{2})u^{2} ],
\end{eqnarray}
\begin{eqnarray}
\gamma_{\phi}(u) = -\epsilon u[ 2b_{2}u + (3b_{3} - 2b_{2}a_{1})u^{2} ],
\end{eqnarray}
\begin{eqnarray}
\overline{\gamma}_{\phi^{2}}(u) = \epsilon u[ c_{1} + (2c_{2} - c_{1}^{2} - 2a_{1}c_{1})u ].
\end{eqnarray}
In turn, the coefficients $a_{1}$, $\cdots$, $c_{2}$ are written in terms of the Feynman diagrams \cite{Amit}. As the definition demands, for obtaining the vertex parts we have to cut their external legs. Thus, for diagrams evaluation purposes, what matters are the internal bubbles of the respective diagrams. So we have to compute the reduced number of diagrams, without taking into account their symmetry factors: \parbox{7mm}{\includegraphics[scale=0.5]{fig10.eps}}($\propto$\hspace{1mm}\parbox{7mm}{\includegraphics[scale=0.5]{fig14.eps}}), \parbox{7mm}{\includegraphics[scale=0.5]{fig6.eps}}, \parbox{7mm}{\includegraphics[scale=0.5]{fig7.eps}} and \parbox{9mm}{\includegraphics[scale=0.5]{fig21.eps}}($\propto$\hspace{1mm}\parbox{8mm}{\includegraphics[scale=0.5]{fig17.eps}}). We also have that \parbox{7mm}{\includegraphics[scale=0.5]{fig11.eps}}\hspace{2mm}$\propto$\hspace{2mm}\parbox{7mm}{\includegraphics[scale=0.5]{fig16.eps}}\hspace{1mm}$\propto$\hspace{2mm}(\parbox{5mm}{\includegraphics[scale=0.5]{fig10.eps}})$^{2}$. Thus by using the results furnished in the \ref{Integral formulas in $d$-dimensional Euclidean momentum space}, \ref{Feynman integrals} and absorbing $\hat{S}$ in a redefinition of $u_{0}$ and $u$ we find 
\begin{eqnarray}\label{uahuahuahu}
\beta(u) = u\left( -\epsilon + \frac{N + 8}{6}\Pi u - \frac{3N + 14}{12}\Pi^{2}u^{2}\right),
\end{eqnarray}
\begin{eqnarray}
\gamma_{\phi}(u) = \frac{N + 2}{72}\left( \Pi^{2}u^{2} - \frac{N + 8}{24}\Pi^{3}u^{3} \right),
\end{eqnarray}
\begin{eqnarray}\label{uahuahuahuaa}
\overline{\gamma}_{\phi^{2}}(u) = \frac{N + 2}{6}\left( \Pi u - \frac{1}{2}\Pi^{2}u^{2} \right).
\end{eqnarray}
where
\begin{eqnarray}
\Pi = 1 - \frac{1}{2}K_{\mu\nu}\delta^{\mu\nu} + \frac{1}{8}K_{\mu\nu}K_{\rho\sigma}\delta^{\{\mu\nu}\delta^{\rho\sigma\}} + ...,
\end{eqnarray}
$\delta^{\{\mu\nu}\delta^{\rho\sigma\}} \equiv \delta^{\mu\nu}\delta^{\rho\sigma} + \delta^{\mu\rho}\delta^{\nu\sigma} + \delta^{\mu\sigma}\delta^{\nu\rho}$ and $\delta^{\mu\nu}$ is the Kronecker delta symbol. As promised by the renormalization program, the $\beta$\hyp{}function, field and composite field anomalous dimensions are finite. Another valuable feature of this renormalization scheme is its elegance and generality, i.e. the theory is renormalized for any values of the external momenta. This fact is expressed by the cancelling of the momentum\hyp{}dependent $L(P)$, $L^{\mu\nu}(P)$, $L_{3}(P)$, $L_{3}^{\mu\nu}(P)$ and $L^{\mu\nu\rho\sigma}(P)$ integrals. This cancelling yields a dependence of the theory on $K$ just through $\Pi$ in a very special functional form: a power-law one. This fact will be used in subsequent developments based on arguments.

\ Now we are in a position to compute the critical exponents. This task is attained by using the so called nontrivial fixed point. Is is obtained from the nontrivial solution to the condition $\beta(u^{*}) = 0$. The trivial one is the Gaussian fixed point, i.e. $u^{*} = 0$. This solution leads to the trivial classical critical exponents in the Landau theory without taking into account the radiative quantum contributions to the exponents. The loop quantum corrections to mean field approximation are furnished by the nontrivial solution    
\begin{eqnarray}\label{yagyaguhd}
u^{*} = \frac{6\epsilon}{(N + 8)\Pi}\left\{ 1 + \epsilon\left[ \frac{3(3N + 14)}{(N + 8)^{2}} \right]\right\}\equiv \frac{u^{(0)}}{\Pi},
\end{eqnarray}
where $u^{(0)}$ is the respective LI nontrivial fixed point \cite{Amit}. By applying the relations $\eta\equiv\gamma_{\phi}(u^{*})$ and $\nu^{-1}\equiv 2 - \eta - \overline{\gamma}_{\phi^{2}}(u^{*})$ we obtain, to next-to-leading order,
\begin{eqnarray}\label{eta}
\eta = \frac{(N + 2)\epsilon^{2}}{2(N + 8)^{2}}\left\{ 1 + \epsilon\left[ \frac{6(3N + 14)}{(N + 8)^{2}} -\frac{1}{4} \right]\right\}\equiv\eta^{(0)},
\end{eqnarray}
\begin{eqnarray}\label{nu}
&&\nu = \frac{1}{2} + \frac{(N + 2)\epsilon}{4(N + 8)} +  \frac{(N + 2)(N^{2} + 23N + 60)\epsilon^{2}}{8(N + 8)^{3}}\nonumber \\ &&\equiv\nu^{(0)},
\end{eqnarray}
where $\eta^{(0)}$ and $\nu^{(0)}$ are the corresponding LI critical exponents \cite{Wilson197475}. Now we proceed to show how this result can be generalized to all-loop level.

\section{All-loop critical exponents in the minimal subtraction scheme}\label{All-loop critical exponents in the minimal subtraction scheme}

\par In Sect. \ref{Minimal subtraction scheme for the critical theory}, for next-to-leading order, in a mathematical viewpoint, although the $\beta$-function and the anomalous dimensions \ref{uahuahuahu}\hyp{}\ref{uahuahuahuaa} present LV corrections, their LV dependence on $\Pi$ (and just on $\Pi$ after it was explicitly shown the cancelling of the momentum-dependent integrals) in a power-law form combined with the one for the nontrivial fixed point \ref{yagyaguhd} yields a cancelling of the LV $\Pi$ factor in the final expressions for the critical exponents. In fact, it was shown earlier \cite{PhysRevD.84.065030} that by applying coordinates transformation techniques, the LV coefficients $K_{\mu\nu}$ can be absorbed into a transformed LI scalar theory with an effective coupling constant given by $u^{(0)} \equiv \Pi u$. This result indicates that the LV theory can be reached if we know the LI one, since we include the $\Pi$ factor in the LV effective coupling constant \cite{PhysRevD.84.065030,Carvalho2013850,Carvalho2014320}. This permits us writing the $\beta$\hyp{}function and anomalous dimensions to any loop-level as
\begin{eqnarray}\label{uhgufhduhufdhu}
\beta(u) = u\left[ -\epsilon + \sum_{n=2}^{\infty}\beta_{n}^{(0)}\Pi^{n-1}u^{n-1}\right], 
\end{eqnarray}
\begin{eqnarray}
\gamma(u) = \sum_{n=2}^{\infty}\gamma_{n}^{(0)}\Pi^{n}u^{n},
\end{eqnarray}
\begin{eqnarray}
\overline{\gamma}_{\phi^{2}}(u) = \sum_{n=1}^{\infty}\overline{\gamma}_{\phi^{2}, n}^{(0)}\Pi^{n}u^{n}.
\end{eqnarray}
where $\beta_{n}^{(0)}$, $\gamma_{n}^{(0)}$ and $\gamma_{\phi^{2},n}^{(0)}$ are the nth\hyp{}loop quantum radiative corrections to the corresponding functions. The Gaussian fixed point is, as it is known, the trivial solution $u^{*}=0$. Moreover the nontrivial fixed point can be evaluated from the nontrivial solution from the bracket in the Eq. \ref{uhgufhduhufdhu}. It is $u^{*} = u^{*(0)}/\Pi$, where $ u^{*(0)}$ is the LI nontrivial fixed point for all\hyp{}loop level. As the $\beta$\hyp{}function and the anomalous dimensions also assume a power-law functional form as in the next-to-leading order approximation, we have again the cancelling of the LV $\Pi$ factor and we finally find that the all-loop order critical exponents are identical to their corresponding LI counterparts. The mathematical explanation for this general result is very helpful for a better comprehension of the problem but is lacked of physical insight. A complete understanding of this result emerges only when we present its physical interpretation. For that, we have to use one of the most solid concepts in the research branch of phase transitions and critical phenomena, i.e. the concept of universality class, discussed in the Sect. \ref{Introduction}. The Lorentz symmetry breaking mechanism occurs in the spacetime in which the fluctuating field is defined, not in its internal symmetry space. So the breaking of this symmetry does not affect the universal values for the critical exponents. In fact, although the LV composite operator is a relevant operator by an initially naive analysis from power-counting considerations, the effect of an operator has to be checked up to the end of the renormalization procedure. Thus this operator does not adds LV contributions to the critical exponents.

\section{Conclusions}\label{Conclusions}

\par In this work we have evaluated, explicitly up to next\hyp{}to\hyp{}leading order, the critical exponents for O($N$) $\lambda\phi^{4}$ scalar field theories with Lorentz violation in the massless regime. We have also computed the same exponents generalizing the results for an arbitrary loop level. For that, we have applied renormalization group and $\epsilon$\hyp{}expansion techniques in the so called massless minimal subtraction scheme where there is no reference of counterterms diagrams and thus a minimal set of Feynman diagrams is needed. We have showed that although the renormalization constants, $\beta$\hyp{}function, anomalous dimensions and fixed point acquire LV corrections, the critical exponents display their universal character as being equal to the LI ones. An identical result has been achieved previously for distinct theories renormalized at different renormalization schemes, showing one more time the universality of the critical exponents.

\appendix
\section{Integral formulas in $d$-dimensional Euclidean momentum space}\label{Integral formulas in $d$-dimensional Euclidean momentum space}

\par Considering $\hat{S}_{d} \equiv S_{d}/(2\pi)^{d} = [2^{d-1}\pi^{d/2}\Gamma(d/2)]^{-1}$ where $S_{d} = 2\pi^{d/2}/\Gamma(d/2)$ is the unit $d$\hyp{}dimensional sphere area, we have

\begin{eqnarray}
&&\int \frac{d^{d}q}{(2\pi)^{d}} \frac{1}{(q^{2} + 2pq + M^{2})^{\alpha}} =  \nonumber \\&&   \hat{S}_{d}\frac{1}{2}\frac{\Gamma(d/2)}{\Gamma(\alpha)}\frac{\Gamma(\alpha - d/2)}{(M^{2} - p^{2})^{\alpha - d/2}},
\end{eqnarray}

\begin{eqnarray}
&&\int \frac{d^{d}q}{(2\pi)^{d}} \frac{q^{\mu}}{(q^{2} + 2pq + M^{2})^{\alpha}} =   \nonumber \\&&  -\hat{S}_{d}\frac{1}{2}\frac{\Gamma(d/2)}{\Gamma(\alpha)}\frac{p^{\mu}\Gamma(\alpha - d/2)}{(M^{2} - p^{2})^{\alpha - d/2}},
\end{eqnarray}

\begin{eqnarray}
&&\int \frac{d^{d}q}{(2\pi)^{d}} \frac{q^{\mu}q^{\nu}}{(q^{2} + 2pq + M^{2})^{\alpha}} = \hat{S}_{d}\frac{1}{2}\frac{\Gamma(d/2)}{\Gamma(\alpha)}   \nonumber \\&&   \times\left[ \frac{1}{2}\delta^{\mu\nu}\frac{\Gamma(\alpha - 1 - d/2)}{(M^{2} - p^{2})^{\alpha - 1 - d/2}}  \right.  \nonumber \\  && + \left. p^{\mu}p^{\nu}\frac{\Gamma(\alpha - d/2)}{(M^{2} - p^{2})^{\alpha - d/2}} \right],
\end{eqnarray}

\begin{eqnarray}
&&\int \frac{d^{d}q}{(2\pi)^{d}} \frac{q^{\mu}q^{\nu}q^{\rho}}{(q^{2} + 2pq + M^{2})^{\alpha}} = - \hat{S}_{d}\frac{1}{2}\frac{\Gamma(d/2)}{\Gamma(\alpha)}  \nonumber \\&&   \times\left[\frac{1}{2}[\delta^{\mu\nu}p^{\rho} + \delta^{\mu\rho}p^{\nu} + \delta^{\nu\rho}p^{\mu}]\frac{\Gamma(\alpha - 1 - d/2)}{(M^{2} - p^{2})^{\alpha - 1 - d/2}}  \right.  \nonumber \\  &&\left. + p^{\mu}p^{\nu}p^{\rho}\frac{\Gamma(\alpha - d/2)}{(M^{2} - p^{2})^{\alpha - d/2}} \right],
\end{eqnarray}

\begin{eqnarray}
&&\int \frac{d^{d}q}{(2\pi)^{d}} \frac{q^{\mu}q^{\nu}q^{\rho}q^{\sigma}}{(q^{2} + 2pq + M^{2})^{\alpha}} = \hat{S}_{d}\frac{1}{2}\frac{\Gamma(d/2)}{\Gamma(\alpha)}  \nonumber \\&&  \times\left[\frac{1}{4}[\delta^{\mu\nu}\delta^{\rho\sigma} + \delta^{\mu\rho}\delta^{\nu\sigma} +\delta^{\mu\sigma}\delta^{\nu\rho}]\frac{\Gamma(\alpha - 2 - d/2)}{(M^{2} - p^{2})^{\alpha - 2 - d/2}}  \right.  \nonumber \\  &&\left. + \frac{1}{2}[\delta^{\mu\nu}p^{\rho}p^{\sigma} + \delta^{\mu\rho}p^{\nu}p^{\sigma} + \delta^{\mu\sigma}p^{\nu}p^{\rho}  \right.  \nonumber \\&&  \left. + \delta^{\nu\rho}p^{\mu}p^{\sigma} +\delta^{\nu\sigma}p^{\mu}p^{\rho} +\delta^{\rho\sigma}p^{\mu}p^{\nu}]\frac{\Gamma(\alpha - 1 - d/2)}{(M^{2} - p^{2})^{\alpha - 1 - d/2}}   \right.  \nonumber \\  &&\left. + p^{\mu}p^{\nu}p^{\rho}p^{\sigma}\frac{\Gamma(\alpha - d/2)}{(M^{2} - p^{2})^{\alpha - d/2}} \right].
\end{eqnarray}

\section{Feynman integrals}\label{Feynman integrals}

\par The diagrams used here are

\begin{eqnarray}
&&\parbox{12mm}{\includegraphics[scale=1.0]{fig10.eps}} = \int \frac{d^{d}q}{(2\pi)^{d}}\frac{1}{q^{2} + K_{\mu\nu}q^{\mu}q^{\nu}} \nonumber \\&&  \times\frac{1}{(q + P)^{2} + K_{\mu\nu}(q + P)^{\mu}(q + P)^{\nu}},
\end{eqnarray}   
\begin{eqnarray}
&&\parbox{12mm}{\includegraphics[scale=1.0]{fig6.eps}} = \int \frac{d^{d}q_{1}}{(2\pi)^{d}}\frac{d^{d}q_{2}}{(2\pi)^{d}}\frac{1}{q_{1}^2 + K_{\mu\nu}q_{1}^{\mu}q_{1}^{\nu}}  \nonumber \\&&  \times\frac{1}{q_{2}^2 + K_{\mu\nu}q_{2}^{\mu}q_{2}^{\nu}}  \nonumber \\&&  \frac{1}{(q_{1} + q_{2} + P)^2 + K_{\mu\nu}(q_{1} + q_{2} + P)^{\mu}(q_{1} + q_{2} + P)^{\nu}}, \nonumber \\
\end{eqnarray}   
\begin{eqnarray}
&&\parbox{12mm}{\includegraphics[scale=1.0]{fig7.eps}} = \int \frac{d^{d}q_{1}}{(2\pi)^{d}}\frac{d^{d}q_{2}}{(2\pi)^{d}}\frac{d^{d}q_{3}}{(2\pi)^{d}}\frac{1}{q_{1}^2 + K_{\mu\nu}q_{1}^{\mu}q_{1}^{\nu}}  \nonumber \\&&  \times\frac{1}{q_{2}^2 + K_{\mu\nu}q_{2}^{\mu}q_{2}^{\nu}}\frac{1}{q_{3}^2 + K_{\mu\nu}q_{3}^{\mu}q_{3}^{\nu}} \nonumber \\ &&\times\frac{1}{(q_{1} + q_{2} + P)^2 + K_{\mu\nu}(q_{1} + q_{2} + P)^{\mu}(q_{1} + q_{2} + P)^{\nu}} \nonumber \\ &&\frac{1}{(q_{1} + q_{3} + P)^2 + K_{\mu\nu}(q_{1} + q_{3} + P)^{\mu}(q_{1} + q_{3} + P)^{\nu}}, \nonumber \\
\end{eqnarray} 
\begin{eqnarray}
&&\parbox{14mm}{\includegraphics[scale=1.0]{fig21.eps}} = \int \frac{d^{d}q_{1}}{(2\pi)^{d}}\frac{d^{d}q_{2}}{(2\pi)^{d}}\frac{1}{q_{1}^2 + K_{\mu\nu}q_{1}^{\mu}q_{1}^{\nu}} \nonumber \\ && \times\frac{1}{(P - q_{1})^{2} + K_{\mu\nu}(P - q_{1})^{\mu}(P - q_{1})^{\nu}} \nonumber \\ && \times\frac{1}{q_{2}^2 + K_{\mu\nu}q_{2}^{\mu}q_{2}^{\nu}} \nonumber \\ && \times\frac{1}{(q_{1} - q_{2} + P_{3})^2 + K_{\mu\nu}(q_{1} - q_{2} + P_{3})^{\mu}(q_{1} - q_{2} + P_{3})^{\nu}}. \nonumber \\
\end{eqnarray}     

\par By making use of the expansion for the free propagator
\begin{eqnarray}\label{expansion}
&& \frac{1}{(q^{2} + K_{\mu\nu}q^{\mu}q^{\nu} + m^{2})^{n}} = \frac{1}{(q^{2} + m^{2})^{n}}\left[ 1 - n\frac{K_{\mu\nu}q^{\mu}q^{\nu}}{q^{2} + m^{2}} \right.  \nonumber \\  &&\left. +  \frac{n(n+1)}{2!}\frac{K_{\mu\nu}K_{\rho\sigma}q^{\mu}q^{\nu}q^{\rho}q^{\sigma}}{(q^{2} + m^{2})^{2}} + ...\right]
\end{eqnarray}
in the small parameters $K_{\mu\nu}$ and the formulas in the \ref{Integral formulas in $d$-dimensional Euclidean momentum space} combined with dimensional regularization and $\epsilon$\hyp{}expansion techniques in $\epsilon = 4 - d$ we have the results
\begin{eqnarray}
&&\parbox{10mm}{\includegraphics[scale=1.0]{fig10.eps}} = \frac{1}{\epsilon}\left\{ \left[1 - \frac{1}{2}\epsilon - \frac{1}{2}\epsilon L(P) \right]\Pi -\frac{1}{2}\epsilon K_{\mu\nu}L^{\mu\nu}(P) \right.  \nonumber \\  &&\left. + \frac{1}{4}\epsilon K_{\mu\nu}K_{\rho\sigma}[L^{\mu\nu}(P)\delta^{\rho\sigma} + L^{\mu\nu\rho\sigma}(P)] \right\},
\end{eqnarray}   
\begin{eqnarray}
&&\parbox{12mm}{\includegraphics[scale=1.0]{fig6.eps}} = -\frac{P^{2}}{8\epsilon}\left[ 1 + \frac{1}{4}\epsilon -2\epsilon L_{3}(P) \right]\Pi^{2} \nonumber \\  && + \frac{P^{2}}{4}K_{\mu\nu}L_{3}^{\mu\nu}(P),
\end{eqnarray}  
\begin{eqnarray}
&&\parbox{10mm}{\includegraphics[scale=0.9]{fig7.eps}} = -\frac{P^{2}}{6\epsilon^{2}}\left[ 1 + \frac{1}{2}\epsilon -3\epsilon L_{3}(P) \right]\Pi^{3} \nonumber \\  && + \frac{P^{2}}{2\epsilon}K_{\mu\nu}L_{3}^{\mu\nu}(P),
\end{eqnarray}  
\begin{eqnarray}
&&\parbox{12mm}{\includegraphics[scale=0.8]{fig21.eps}} = \frac{1}{2\epsilon^{2}}\left\{ \left[1 - \frac{1}{2}\epsilon - \epsilon L(P) \right]\Pi^{2} \right.  \nonumber \\  &&\left. -\epsilon K_{\mu\nu}L^{\mu\nu}(P) \right.  \nonumber \\  &&\left. + \frac{1}{2}\epsilon K_{\mu\nu}K_{\rho\sigma}[L^{\mu\nu}(P)\delta^{\rho\sigma} + L^{\mu\nu\rho\sigma}(P)] \right\},
\end{eqnarray}  
where
\begin{eqnarray}
&&L(P) = \int_{0}^{1}dx\ln[x(1-x)P^{2}],
\end{eqnarray}
\begin{eqnarray}
&&L^{\mu\nu}(P) = \int_{0}^{1}dx\frac{x(1-x)P^{\mu}P^{\nu}}{x(1-x)P^{2}},
\end{eqnarray}
\begin{eqnarray}
&&L_{3}(P) = \int_{0}^{1}dx(1-x)\ln[x(1-x)P^{2}],
\end{eqnarray}
\begin{eqnarray}
&&L_{3}^{\mu\nu}(P) = \int_{0}^{1}dx\frac{x(1-x)^{2}P^{\mu}P^{\nu}}{x(1-x)P^{2}},
\end{eqnarray}
\begin{eqnarray}
&&L^{\mu\nu\rho\sigma}(P) = \int_{0}^{1}dx\frac{x^{2}(1-x)^{2}P^{\mu}P^{\nu}P^{\rho}P^{\sigma}}{[x(1-x)P^{2}]^{2}}.
\end{eqnarray}
The diagrams $\parbox{5mm}{\includegraphics[scale=.5]{fig10.eps}}$ and $\parbox{7mm}{\includegraphics[scale=.5]{fig21.eps}}$ are evaluated perturbatively in the small parameters $K_{\mu\nu}$ up
to $\mathcal{O}(K^{2})$ and $\parbox{6mm}{\includegraphics[scale=.5]{fig6.eps}}$ and $\parbox{6mm}{\includegraphics[scale=.5]{fig7.eps}}$ up to $\mathcal{O}(K)$. The computation of the latter couple of diagrams up to $\mathcal{O}(K^{2})$ would be very tedious \cite{Carvalho2014320}.


\begin{acknowledgements}
William C. Vieira would like to thank the Brazilian funding agency CAPES for financial support.
\end{acknowledgements}

\bibliography{apstemplate}

\end{document}